\documentclass[reprint,aps,prl,twocolumn,superscriptaddress,showkeys,showpacs,
longbibliography,groupedaddress]{revtex4-1}
\usepackage[utf8]{inputenc}
\usepackage{graphicx}
\usepackage{amssymb, amsmath}
\usepackage{color,soul}
\usepackage{float}
\usepackage{hyperref}
\usepackage{calc}

\begin{document}

\title{Phase ordering kinetics of the long-range Ising model}
\author{Henrik Christiansen}
\email{henrik.christiansen@itp.uni-leipzig.de}
\author{Suman Majumder}
\email{suman.majumder@itp.uni-leipzig.de}
\author{Wolfhard Janke}
\email{wolfhard.janke@itp.uni-leipzig.de}
\affiliation{Institut für Theoretische Physik, Universität Leipzig, Postfach 100 
920, 04009 Leipzig, Germany}
\date{\today}
 
 \begin{abstract}
We use an efficient method that eases the daunting task of simulating dynamics 
in spin systems with 
long-range interaction. Our Monte Carlo simulations of the long-range Ising 
model for the nonequilibrium phase ordering dynamics 
in two spatial dimensions perform significantly faster than the standard Metropolis approach
and considerably more efficiently than the kinetic Monte Carlo method. Importantly, this enables us to establish agreement with the 
theoretical prediction for the time dependence of domain growth,
in contrast to previous numerical studies. This method can easily be generalized 
to applications in other systems.
\end{abstract}

\maketitle
Generic models of statistical physics exhibiting a transition from disordered to 
ordered states
have been proved to be instrumental for understanding the dynamics in diverse 
fields,
from species evolution \cite{Bak} to traffic flow \cite{Bak}, 
from economic dynamics \cite{Economics} to rainfall dynamics \cite{rainfall}.
An extensively used paradigm is the Ising model with nearest-neighbor (NNIM) 
interaction \cite{Bray_article,puri2009kinetics}. 
Even the complex neural dynamics of brain depends on similar underlying 
mechanisms \cite{Beggs}. 
The maximum entropy models obtained from experimental data upon mapping 
the spiking activities of the neurons onto {spin variables} are equivalent to 
Ising models \cite{Schneidman}. 
However, it is believed that the neuron activities are effectively modelled by 
long-distance communications \cite{Beggs}.
In nature, also many other intermolecular interactions are evidently
long-range, e.g., electrostatic forces, polarization forces, etc.
Hence, a more complete picture calls for employing models that consider 
long-range interactions.
\par
The simplest generic model system is the long-range 
Ising model (LRIM), which on a $d$-dimensional lattice is described by the 
Hamiltonian 
\begin{equation}\label{LRIM}
\mathcal{H}=-\sum_{i} \sum_{j < i} J(r_{ij}) 
s_is_j,~\textrm{with}~J(r_{ij})=\frac{1}{r_{ij}^{d+\sigma}},
\end{equation}
where spins $s_i =\pm 1$, $r_{ij}$ is the distance between the 
spins at site $i$ and $j$, and $J(r_{ij})$ is the interaction strength. 
The model exhibits a para- to ferro-magnetic phase transition.
Naturally, simulations of such systems with long-range 
interaction 
are computationally far more expensive than its short-range counterpart. 
For equilibrium studies, the advent of various collective 
updates based on the Swendsen-Wang cluster algorithm \cite{Swendsen-Wang} allows 
one
to perform efficient Monte Carlo (MC) simulations 
\cite{luijten1995monte,fukui2009order,flores2017cluster}.
Conversely, for understanding the nonequilibrium ordering kinetics 
following 
a quench from the high-temperature disordered phase into the ordered phase 
below 
the critical temperature $T_c$, one is restricted to use only local moves, 
viz., 
single spin flips. This makes MC simulations of ordering kinetics in LRIM 
severely expensive even with present-day computational facilities, and 
therefore, they have rarely been 
attempted \cite{gundh2015ordering}. 
\par
The understanding of ferromagnetic ordering kinetics in NNIM is well developed 
\cite{Bray_article,puri2009kinetics}. 
It is characterized by formation and growth of domains of like spins and is a 
scaling phenomenon, i.e., the characteristic length scale a.k.a. 
the domain size $\ell(t)$ at time $t$ follows the Lifshitz-Cahn-Allen (LCA) law 
\cite{Bray_article}: $\ell(t) \sim t^{1/2}$ which can be derived by 
considering that $\ell(t)$ grows via reduction of the curvature $1/\ell(t)$ of 
the domain walls. Similarly for 
the LRIM the growth is likely to be driven by interactions between domain 
walls. 
Assuming this growth as a scaling phenomenon and using an ``energy scaling'' 
argument it has been predicted that 
\cite{Bray_only,bray1994growth,Rutenberg1994}
\begin{equation}
 \ell(t)\propto t^{\alpha}=
\begin{cases}
 t^{\frac{1}{1+\sigma}} & \sigma < 1 \\
 (t \ln t)^{\frac{1}{2}} & \sigma = 1 \\
 t^{\frac{1}{2}} & \sigma > 1
\end{cases},
\label{Prediction}
\end{equation}
i.e., (i) in the ``truly'' long-range regime for $\sigma<1$, the growth exponent 
$\alpha$ is $\sigma$ dependent, (ii) at the crossover point $\sigma=1$, the 
growth follows the LCA law with a multiplicative logarithmic correction, and 
(iii) for $\sigma>1$, LRIM behaves asymptotically as the NNIM with $\alpha=1/2$.
There exist few attempts to confirm these predictions via numerical solution of 
Ginzburg-Landau-type \cite{Hisao93} or Langevin-type \cite{Lee93,Ispolatov1999} 
dynamical equations. The only available results from MC simulations 
\cite{gundh2015ordering} in this regard tackles the expensive calculation of the 
local energy involving all the spins by using a cut-off distance for $J(r_{ij})$ 
in \eqref{LRIM}. Importantly, in disagreement with \eqref{Prediction}, $\alpha$ 
is found there 
to be no different than in NNIM for all $\sigma$, thus suggesting a universal 
nonequilibrium behavior.
In equilibrium it is well established both theoretically 
\cite{Stell1970,Fisher1972,Sak73} and in simulations 
\cite{luijten1997,Luijten2002,horita2017upper} 
that critical exponents are not universal. For example, in the $d=2$ LRIM, for $\sigma < 1$ 
the critical exponent $\eta$ takes its mean-field value, followed by an 
intermediate range $1 < \sigma < \sigma_{\times}$ where it is 
$\sigma$-dependent, and for $\sigma > \sigma_{\times}$ it behaves like in the 
NNIM. 
The value of the crossover point $\sigma_{\times}$ is still disputed 
\cite{horita2017upper} and predicted to be $\sigma_{\times}=2$ \cite{Fisher1972} 
or $\sigma_{\times}=7/4$ \cite{Sak73}.
In this Letter, we present results from MC 
simulations for the ordering kinetics of LRIM in $d=2$ using our efficient approach with the aim 
to check the $\sigma$-dependence of the growth exponent $\alpha$.
\par
In a standard Metropolis simulation \cite{landau2014guide} for kinetics of LRIM 
one attempts to flip a randomly chosen spin $s_i$ with 
probability $p_i=\min[1,\exp(-\Delta E_i/k_BT)]$, where $k_B(=1)$ is the 
Boltzmann constant, $T$ is the temperature 
and $\Delta E_i$ is the change in energy due to the 
flip. The aim of our approach is to avoid the expensive calculation of $\Delta E_i$ at every 
attempt. Instead we store the effective field, 
assigned to each spin, and only update other spin flips to this effective field 
\footnote{We thank A. Hucht for informing us after completion of this work, that such an approach has already been very briefly mentioned in Ref.~\cite{HUCHT199532} where the method was used to simulate the Heisenberg model with dipolar interactions.}.
\begin{table}[!b]
\caption{Average number of POSIX clocks needed by our method for different 
values of $\sigma$. 
Estimations are made from simulations of LRIM with $L=1024$ averaged over $20$ 
initial realizations, running up to $10^4$ MCS. Corresponding clocks for the 
standard method are $5.0(1) \times 10^{13}$. All simulations were run on a 
Intel Xeon 
CPU E5-2640 v4.} 
 \begin{tabular}{cccccc}
  \hline
  \hline
  $\sigma$ & $0.4$ & $0.6$ & $0.8$ & $1.0$ & $1.5$ \\
  \hline
  Clocks ($10^{10}$) & $1.42(4)$ & $2.1(2)$ & $3.7(2)$ & $5.3(3)$ & $8.1(3)$ \\
  \hline
  \hline
 \end{tabular}
\label{RuntimeTable}
\end{table}
\par
When simulating a long-range interacting system using periodic boundary 
conditions (via minimum-image convention), 
one encounters strong finite-size effects. We circumvent this problem by 
using Ewald summation \cite{ewald1921berechnung,horita2017upper,flores2017cluster} for calculating the effective interaction $J(r_{ij})$.
To prepare an initial configuration that mimics a high-temperature paramagnetic phase ($T\gg T_c$)
we choose a square lattice having linear dimension $L$ with randomly $50 \%$ up and $50\%$ down spins. 
Next, for each spin $s_i$, we store the effective field
\begin{equation}\label{heatbath}
h_i=\sum_{j\ne i}J(r_{ij})s_j.
\end{equation}
The Metropolis simulation at any given temperature can now be done efficiently 
with the advantage of having these stored $h_i$, in the following way.
Using Eq.\ \eqref{heatbath} one can write down the change in energy due to an attempted 
flip of a randomly chosen spin $s_i$ as
\begin{equation}\label{energy_change}
\Delta E_i =E_i^{\rm{new}}-E_i^{\rm{old}}=2 s_i\sum_{j\ne i}J(r_{ij})s_j=2s_ih_i.
\end{equation}
Now if the spin $s_i$ is flipped the effective field $h_j$ of 
any other spin $s_j$ accounts for a change of $-2s_iJ(r_{ij})$, thus $h_j \rightarrow h_j-2s_iJ(r_{ij})$. This operation can be performed with roughly the same computational effort as calculating a 
single $\Delta E_i$ in the traditional approach. 
However, one does this only for accepted spin flips.
Thus many spin-flip attempts can be made without this update of $h_j$, facilitating a significant speedup.
\par
The above approach is reminiscent of the $n$-fold way or kinetic MC (KMC) simulations \cite{n-fold,Voter}, which have been extensively used for short-range models.
In KMC for the NNIM the major advantage lies in categorizing the local spin environment into classes. To the best of our knowledge, KMC has never been applied 
in the LRIM, presumably because construction of classes is impossible in the long-range case and the probability of every spin flip needs to be calculated at each step.
Combining the idea of updating the
effective fields or the probabilities during KMC, of course, improves the performance, but even then, our approach provides $\sim 5$ times better performance \footnote{In our approach the average waiting time between spin-flips is $\sim L$, as it can be shown that for the coarsening period the average acceptance $\langle P_{\textrm{acc}} \rangle \sim 1/L$. Thus we have to calculate $p_i$ involving the computationally expensive exponential function, 
only $\sim L$ times, whereas in KMC, the exponential function has to be calculated $L^2$ times always.} at quench temperature $T_q=0.1T_c$ 
\footnote{The value of $T_c(\sigma)$ for the LRIM is 
extracted by a power-law fit of form $T_{c}(\sigma) = T_{c}(\infty) + a 
\sigma^{b} $ to the data presented in Ref. \cite{horita2017upper}. 
For the purpose of nonequilibrium simulations the precise value of $T_c$ is not 
important and this rather crude estimate is sufficient.}, which will be used subsequently for all our simulations. 
For this and all following analyses, the unit of time is one MC step (MCS) that consists of $L^2$ spin-flip attempts. 
The results for the ordering kinetics are averaged over $50$ independent realizations for $L=2048$ and $100$ realizations for $L<2048$.
\begin{figure}[!t]
\centering
\includegraphics{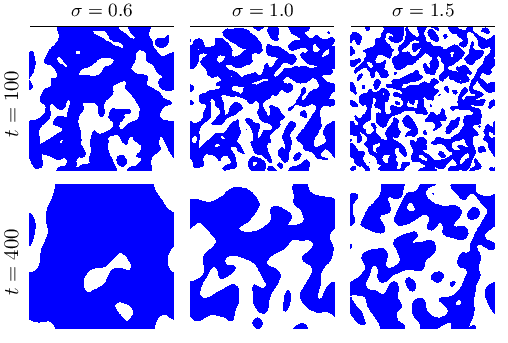}
 \caption{Evolution snapshots at different times, 
demonstrating the ferromagnetic ordering in LRIM with $L=1024$ for different 
$\sigma$. Only the up spins ($+$) are marked.}
 \label{Evolution}
 \end{figure}
 \begin{figure*}[!t]
\centering
 \includegraphics{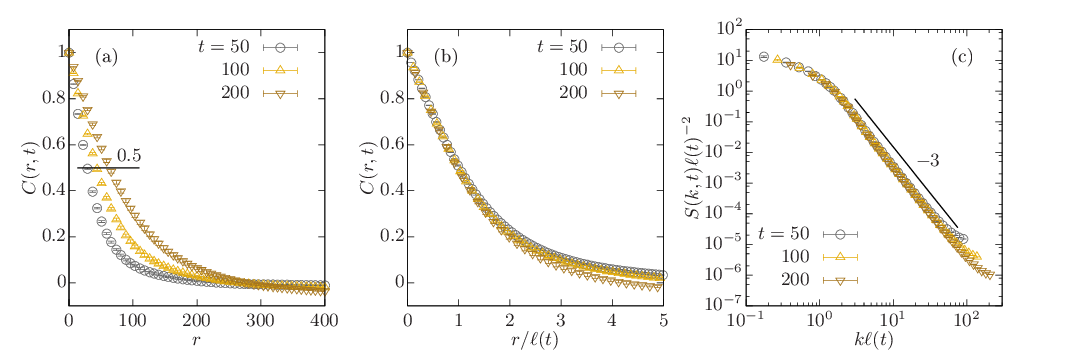}
 \caption{(a) Correlation functions $C(r,t)$ at different times for 
$\sigma=0.6$. (b) Demonstration of the scaling of $C(r,t)$ as a 
function of $r/\ell(t)$ for the same times as in (a). (c) Scaling plots 
for the structure factor  $S(k,t)$. The solid line there corresponds to the 
Porod tail behavior of 
$S(k,t) \sim k^{-3}$.}
 \label{Correlation}
\end{figure*} 
\par
In Table\ \ref{RuntimeTable} we tabulated the number of CPU clocks needed for 
our method to perform $10^4$ MCS for different 
$\sigma$. Roughly the clock time for all the $\sigma$ is $\sim 10^{10}$. To run 
the same number of MCS using the standard approach the 
clock time is $\sim 10^{13}$. Thus an improvement factor $\approx 10^3$ can be 
achieved with this algorithm for the LRIM at the chosen quench temperature.
Since for our method the lower the acceptance rate the more one gains in speed, 
at lower temperatures the efficiency gain with respect to the standard approach 
becomes higher, whereas at $T=\infty$ both of them should have identical run 
time. Note that our algorithm becomes faster as the simulation 
moves on because of the 
lower acceptance rates when the system approaches the ordered phase, and we emphasize it 
does not use any cut-off in $J(r_{ij})$.
\par 
Having the new methodology in place, we move on to explore the kinetics of the 
ferromagnetic ordering in LRIM. In Fig.\ \ref{Evolution} we present evolution 
snapshots for three different values of $\sigma$ 
from a typical quench.
Apparently the structural changes during the evolution are no different than in 
NNIM \cite{Bray_article}. From the snapshots at the same time 
for different $\sigma$ it is evident that the smaller the value of $\sigma$ the 
faster is the growth. However, one needs to estimate the 
growth exponents in order to overrule the claim of the scaling equivalence for 
different $\sigma$ reported in Ref.\ \cite{gundh2015ordering}.

\begin{figure}[!b]
\centering
\includegraphics{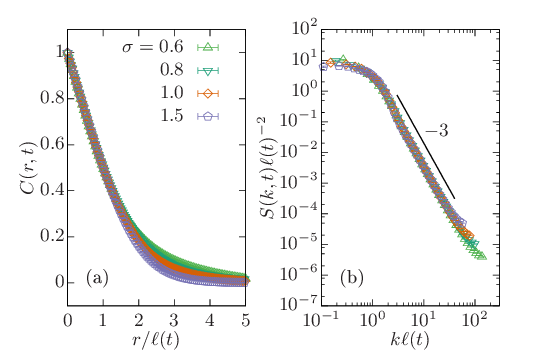}
\caption{(a) Scaled correlation function $C(r,t)$ at $t=100$ MCS for different 
$\sigma$ as mentioned. 
(b) Same as (a) but for the scaled structure factor $S(k,t)$. The solid line 
again corresponds to the Porod tail.}
\label{Structure}
\end{figure}
\begin{figure}[!t]
 \includegraphics{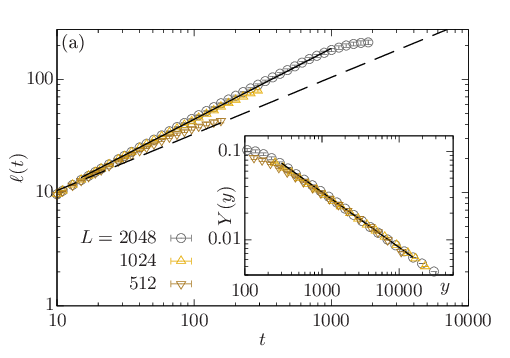}
 \includegraphics{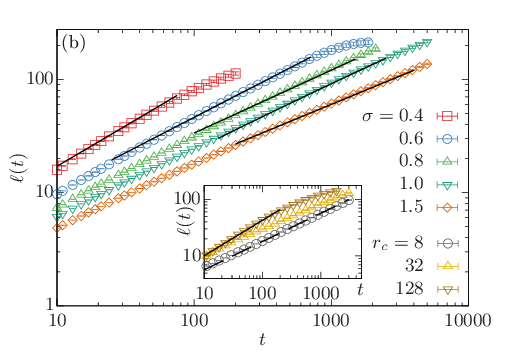}
 \caption{(a) Time dependence of the length scale $\ell(t)$ for 
$\sigma=0.6$ for three different $L$ as indicated.  
 The solid and the dashed lines correspond to $t^{1/(1+\sigma)}$ and $t^{1/2}$, 
respectively. The inset illustrates the finite-size scaling
 using the same data with $t_0=3$ and $\ell_0=5$. The solid line shows the expected $y^{-1/(1+\sigma)}$ behavior.
 (b) Time dependence of $\ell(t)$ for different $\sigma$ as indicated with 
$L=2048$. The solid lines 
 are the respective predictions in \eqref{Prediction}. The inset here shows the length-scale data for $\sigma=0.6$ with $L=1024$ 
 when using different cut-off distances $r_c$ in Eq.\ \eqref{LRIM}. The cut-off 
$r_c=8$ is close to the value of $r_c \approx 8.3$ used in Ref.\ 
\cite{gundh2015ordering}. The lines 
have the same meaning as in the main frame of (a).}
 \label{Length}
\end{figure}
\par
We now check the scaling of the morphology-characterizing 
two-point equal-time correlation function 
$ C(r,t)=\langle s_is_j\rangle - \langle s_i \rangle \langle s_j\rangle$
and its Fourier transform, the structure factor $S(\vec{k},t)=\int 
d\vec{r}C(\vec{r},t)e^{i\vec{k}\vec{r}}$. 
Figure\ \ref{Correlation}(a) presents $C(r,t)$ at different times for 
$\sigma=0.6$, showing 
the signature of a growing length scale with time. 
The multiplicative scaling during the growth is confirmed by the 
data collapse as shown in Fig.\ \ref{Correlation}(b), 
on plotting the $C(r,t)$ against $r/\ell(t)$ where the length scale 
$\ell(t)$ is extracted from the criterion $C[r=\ell(t),t]=0.5$. 
The data at large $r/\ell(t)$ for the latest time seems to show some discrepancy 
attributed to finite-size effects. 
However, the scaling of the structure factor 
$S(\vec{k},t)$ that forms a basic 
assumption when deriving the theoretical growth laws for LRIM 
\cite{bray1994growth,Rutenberg1994}, is confirmed convincingly as shown in Fig.\ 
\ref{Correlation}(c). 
Similar respective behavior is observed when scaled $C(r,t)$ and $S(k,t)$ at the 
same time are plotted for different $\sigma$ in Fig.\ \ref{Structure}. 
The slower decay of $C(r,t)$ for smaller values of $\sigma$ 
could be an indication of the inverse relation of the 
growth exponent $\alpha$ with $\sigma$, as predicted in 
Eq.\ \eqref{Prediction}. Contrasting, the scaled $S(k,t)$ for different 
$\sigma$ in Fig.\ \ref{Structure}(b) show reasonably good overlap. The solid 
lines in Fig.\ \ref{Correlation}(c) and Fig.\ \ref{Structure}(b) depict the 
consistency of the data with the Porod tail \cite{glatter1982small}: $S(k,t) 
\sim k^{-(d+1)}$ at large wave number $k$. 
\par
The multiplicative scaling of the morphology-characterizing functions indeed 
suggests the presence of scaling of the growing {length scale}. 
Hence, shifting our focus on the growth exponent $\alpha$ in Fig.\ 
\ref{Length}(a) we present the time dependence of the length scale $\ell(t)$ 
for $\sigma=0.6$. The behavior is certainly not $\sim t^{1/2}$ 
(shown by the dashed line), but in fact the data for all $L$ follow the 
predicted behavior of $t^{1/(1+\sigma)}$ until 
they show deviations due to finite-size effects. 
This already indicates that the underlying scaling behavior is indeed consistent 
with \eqref{Prediction}.
Nevertheless, to further strengthen the claim and 
to gauge the effect of a finite system size we call for a finite-size scaling 
(FSS) analysis \cite{Majumder2010,das2012finite} which recently 
has been successfully employed in kinetics of other 
systems \cite{majumder2015cluster,majumder2017kinetics,christiansen2017coarsening}. 
Quantifying the growth including an initial crossover time $t_0$ and length 
$\ell_0=\ell(t_0)$ one can write down the 
ansatz $\ell(t)=\ell_0+A(t-t_0)^{\alpha}$ and construct a FSS function 
$Y(y)=(\ell(t)-\ell_0)/(L-\ell_0)$ with the scaling variable 
$y=(L-\ell_0)^{1/\alpha}/(t-t_0)$. 
In the scaling regime one expects $Y\sim y^{-\alpha}$. Thus on plotting $Y$ as a 
function of $y$ for 
different $L$ one must observe a data collapse with $Y\sim y^{-\alpha}$ behavior 
for large $y$ provided $\alpha$ is chosen appropriately. 
We did this exercise for different $\sigma$ choosing $\alpha$ from 
\eqref{Prediction}. 
However, not all of them are presented here, but rather a representative plot 
for $\sigma=0.6$ is shown in the inset of Fig.\ \ref{Length}(a). The collapsed 
data is 
consistent with the underlying master curve $Y\sim y^{-\alpha}$. 
Considering the collapsed data for all $L$ and fitting 
the ansatz $Y \sim y^{-\alpha}$ by treating 
$\alpha$ ($=1/(1+\sigma)$) as a fit parameter, we obtain $\sigma=0.605(4)$ with 
reasonable reduced chi-squared 
$\chi^2_r=3.47$ within the range $y \in [300,10^4]$.
Similarly, if we fix $\alpha=1/1.6$ according to \eqref{Prediction} and use the 
same fit range as above 
we again get a reasonable $\chi^2_r=4.36$.
\par
In Fig.\ \ref{Length}(b) we present the time dependence of the length scale 
$\ell(t)$ for different 
$\sigma$. Our data clearly indicates that $\alpha$ becomes larger as $\sigma$ decreases. 
Importantly 
in each case the data follows the theoretically predicted behavior 
\eqref{Prediction} shown as solid lines, in contradiction with results 
\cite{gundh2015ordering} 
reporting $\alpha=1/2$ independent of $\sigma$. 
At the crossover point $\sigma=1$ our data follows the LCA growth with 
multiplicative logarithmic correction: $(t \ln t)^{\frac{1}{2}}$, albeit a 
power-law growth with $\alpha>1/2$ cannot unprejudiced be ruled out.
However, in accordance with \eqref{Prediction} for $\sigma=1.5$ in the 
post-crossover regime ($\sigma>1$) the growth appears to be $\sim t^{1/2}$, as 
expected for the NNIM. To consolidate the visual validation we also performed 
for 
each case least-square fits of prediction \eqref{Prediction} and verified the 
predicted exponent values \cite{christiansen2018unpublished}.
In the inset of Fig.\ \ref{Length}(b) we show a plot of the
length scale obtained from simulations using different cut-off radii $r_c$ 
in Eq.\ \eqref{LRIM} for $\sigma=0.6$. 
For the largest $r_c$ the data follows $\sim t^{1/(1+\sigma)}$ behavior as is 
observed without any cut-off, whereas the cases with smaller $r_c$ obey the LCA 
law. 
Thus, in conjunction with the previously reported simulation 
\cite{gundh2015ordering} 
one can infer that the use of a relatively small $r_c$ make the spins interact 
only on short range leading to $\sigma$-independent growth exponents.
\par
To conclude, we have studied the kinetics of ferromagnetic ordering using the 
long-range Ising model in $d=2$ spatial dimensions via MC simulations 
using an efficient method. We have introduced the idea of storing the effective field for each spin that helps to reduce the expensive 
calculation of 
local energy changes involving all the spins at every step. Our 
approach speeds up the simulation by a factor of $\sim 10^3$ 
compared to the standard Metropolis algorithm, and is even considerably faster than the 
efficient kinetic MC method. This enables us to simulate systems as 
big as $2048^2$ spins without using any cut-off radius 
in 
the distance-dependent power-law interaction. Results obtained from our 
simulations are the first confirmation of the theoretical prediction in 
\eqref{Prediction} for the 
growth laws in the long-range Ising model 
\cite{bray1994growth,Bray_only,Rutenberg1994}. We have also demonstrated 
that the inappropriate use 
of a cut-off radius in the local-energy calculation may lead to a different 
growth exponent, explaining the mismatch between previous simulation 
results \cite{gundh2015ordering} and theory.
\par
In equilibrium, the long-range Ising model has a dimension-dependent crossover 
behavior of the critical exponents 
\cite{Stell1970,Fisher1972,Sak73,luijten1997,Luijten2002,horita2017upper}, while 
in nonequilibrium the prediction \eqref{Prediction} is expected to be 
independent of the dimension. In this light, we take the ordering kinetics of 
the $d=3$ case as our next endeavor to check this dimension independence 
\footnote{Work in progress}.
Our method shall trigger interests to explore other aspects 
associated with ordering phenomena in the long-range Ising model, 
viz., 
aging and related dynamical scaling \cite{henkel2010non}. The generic simple 
feature of the method shall ensure its facile adoptions to nonequilibrium 
simulations of other
models, viz., $q$-state Potts and clock models. In view of the delicate cut-off 
dependence, it would also be interesting to revisit the ordering 
phenomenon in long-range liquid crystals \cite{singh2014ordering}. Although 
originally designed for simulating dynamics, our method should be 
proven to be handy for equilibrium simulations of systems with long-range 
interactions, for which there (currently) exist no cluster algorithms, e.g., 
\mbox{(lattice-)} polymers \cite{Note3}.

\begin{acknowledgments}
This project was funded by the Deutsche Forschungsgemeinschaft (DFG) under Grant 
Nos.\ JA 483/33-1 and SFB/TRR 102 (project B04), 
and further supported by the Leipzig Graduate School of Natural Sciences 
``BuildMoNa'', 
the Deutsch-Französische Hochschule (DFH-UFA) through the Doctoral College 
``$\mathbb{L}^4$'' 
under Grant No.\ CDFA-02-07, and the EU Marie Curie IRSES network DIONICOS under 
Grant No.\ PIRSES-GA-2013-612707. We 
thank Stefan Schnabel for a very important suggestion.
\end{acknowledgments}

 \end{document}